\documentclass{emulateapj}
\usepackage{apjfonts}

\newcommand{\Msun}{$\mathrm{M_\odot}$}

\shorttitle{Evolution of massive stars with pulsation-driven superwinds during the RSG phase}
\shortauthors{Yoon \& Cantiello}
 
\usepackage{amsmath} 
\begin{document} 
 
 
\title{Evolution of Massive Stars with Pulsation-Driven Super-Winds during the Red Supergiant Phase}

 
\author{Sung-Chul Yoon \altaffilmark{1, 2} and Matteo Cantiello \altaffilmark{2}} 
 
 
 
\altaffiltext{1}{UCO/Lick Observatory, Department of Astronomy and Astrophysics, University of 
  California, Santa Cruz, CA 95064, USA} 
\altaffiltext{2}{Argelander Insitute for Astronomy, Univeristy of Bonn, Auf dem Huegel 71, D-53121, Germany, scyoon@astro.uni-bonn.de, cantiello@astro.uni-bonn.de}
 
 
\begin{abstract} 

Pulsations driven by partial ionization of hydrogen in the envelope are often
considered important for driving winds from red supergiants (RSGs). In
particular, it has been suggested by some authors that the pulsation growth
rate in a RSG can be high enough  to trigger an unusually strong wind (or a
super-wind), when the luminosity to mass ratio becomes sufficiently large. Using
both hydrostatic and hydrodynamic stellar evolution models with initial masses
ranging from 15 to 40~\Msun, we investigate 1) how the pulsation growth rate
depends on the global parameters of supergiant stars, and 2) what would be
the consequences of a pulsation-driven super-wind, if it occurred, for the late
stages of massive star evolution. We suggest that such a super-wind history
would be marked by a runaway increase, followed by a sudden decrease, of the
winds mass loss rate.  The impact on the late evolution of massive stars would
be substantial, with stars losing a huge fraction of their H-envelope even with
a significantly lower initial mass than previously predicted.  This might
explain the observed lack of Type II-P supernova progenitors having initial
mass higher than about 17~\Msun.  We also discuss possible implications for a
subset of Type IIn supernovae. 
 
\end{abstract}

\keywords{stars: evolution, stars: mass-loss,  stars: massive, supergiants, supernovae:general} 
 
\section{Introduction} 

Mass loss due to stellar winds is one of the governing factors that determine
the evolution of massive stars.  Great improvements in our understanding of
line-driven winds from massive hot stars (i.e., O-Type stars or Wolf-Rayet
stars) have been achieved over the last three decades, and they are well
reflected in recent stellar evolution models \citep[see][for a recent review]{Puls08}.  However,
the driving mechanism for  cool-star winds  remains elusive. 
Note that many of
the authors still rely on the empirical mass-loss rate given by
\citet{deJager88} (hereafter, JNH88), which is more than 20 years old, for
modeling cool giant stars, while some alternative prescriptions have been considered
by some authors \citep[e.g.][]{Salasnich99, vanLoon05, vanBeveren07}.
Late stages of massive star evolution during
and after core helium burning are therefore relatively difficult to study, 
which is a major obstacle to theoretical investigation on 
supernova progenitors and their environments.

One of the key factors that should be taken into account for the study of the
red supergiant (RSG) evolution is pulsation. It is well known that RSGs are
unstable to radial pulsations driven by partial ionization of hydrogen in the
outermost layers of the envelope \citep[e.g.][]{Li94}, and pulsation is one of
the most commonly invoked mechanisms for driving winds from cool stars
\citep[cf.][]{Bowen88, Hoefner03, Neilson08}.  Although the detailed mechanisms
for driving mass loss by pulsations are not well understood,  simple arguments
based on the gain of pulsation energy suggest that RSGs might experience a
stronger mass loss for a higher growth rate of the pulsational instability
\citep[e.g.][]{Appenzeller70}. In particular, \citet{Heger97} found that RSG
pulsations can achieve large amplitudes that may be comparable to those in AGB
stars when the pulsation periods become close to the thermal time scale of the
envelope. They therefore suggested that such strong pulsations might lead to a
"superwind" if the luminosity to mass ratio in a RSG were sufficiently high.
This idea might be supported by observations. 
For example, \citet{vanLoon05} found that the dust-enshrouded RSGs in the Large Magellanic
Cloud have a very high mass loss rate, which appears to be 
related to strong RSG pulsation~\citep{vanLoon08}.

In the present study, we explore possible consequences of such a superwind in
the evolution of supernova progenitors. 
Our approach to this topic is as
follows. First, we investigate the properties of RSG pulsations  using our new
stellar evolution models that follow the non-linear evolution of pulsation
(Sect.~\ref{sect:properties}).  This allows us to derive a relation between the
pulsation growth rate and the structure of RSGs. Second, we assume that the
mass loss rate due to RSG winds should be significantly enhanced compared to
that of JNH88, when the growth rate of pulsation becomes sufficiently large.
The consequent evolution of massive stars can be very different from the
standard picture given in previous work (Sect.~\ref{sect:evol}).  
Note that here we apply the modified mass loss prescription only to very unstable RSGs, 
while \citet{Salasnich99} and \citet{vanBeveren07} considered 
overall enhancement of RSG mass loss rate for all types of RSGs. 
Also, our mass loss prescription is more directly related to the pulsational properties
of RSGs than the others that rely on surface luminosity and temperature. 
Finally, we conclude the paper by discussing implications for supernova progenitors
(Sect.~\ref{sect:discussion}).

\section{Properties of RSG pulsations}\label{sect:properties}

RSG pulsations are not usually found in stellar evolution models even if one
uses a hydrodynamic code, for two reasons.  First, in most stellar evolution
calculations, the evolutionary time steps are much larger than the pulsation
periods in RSGs, which are comparable to the dynamical time scale of the
envelope.  Second, pulsations tend to be numerically damped out with the
implicit method adopted in stellar evolution codes.  However, the study by
\citet{Heger97} shows that the predictions of linear stability analysis for
basic pulsational properties (i.e., pulsation period and growth rate) can be
accurately reproduced in non-linear evolutionary calculations of RSGs by a
hydrodynamic stellar evolution code, if the growth rate is sufficiently large
and if the adopted time steps are sufficiently small.  On the other hand,
non-linear evolutionary calculations do not require the condition of complete
thermal and hydrostatic equilibrium of the star, in contrast to the case of
linear stability analysis.  
We therefore decide to use an
existing hydrodynamic stellar evolution code for our analysis of RSG
pulsations.  Details of our implicit hydrodynamic code are described in
\citet{Yoon06} and references therein. 

\begin{figure}
\epsscale{1.0} 
\plotone{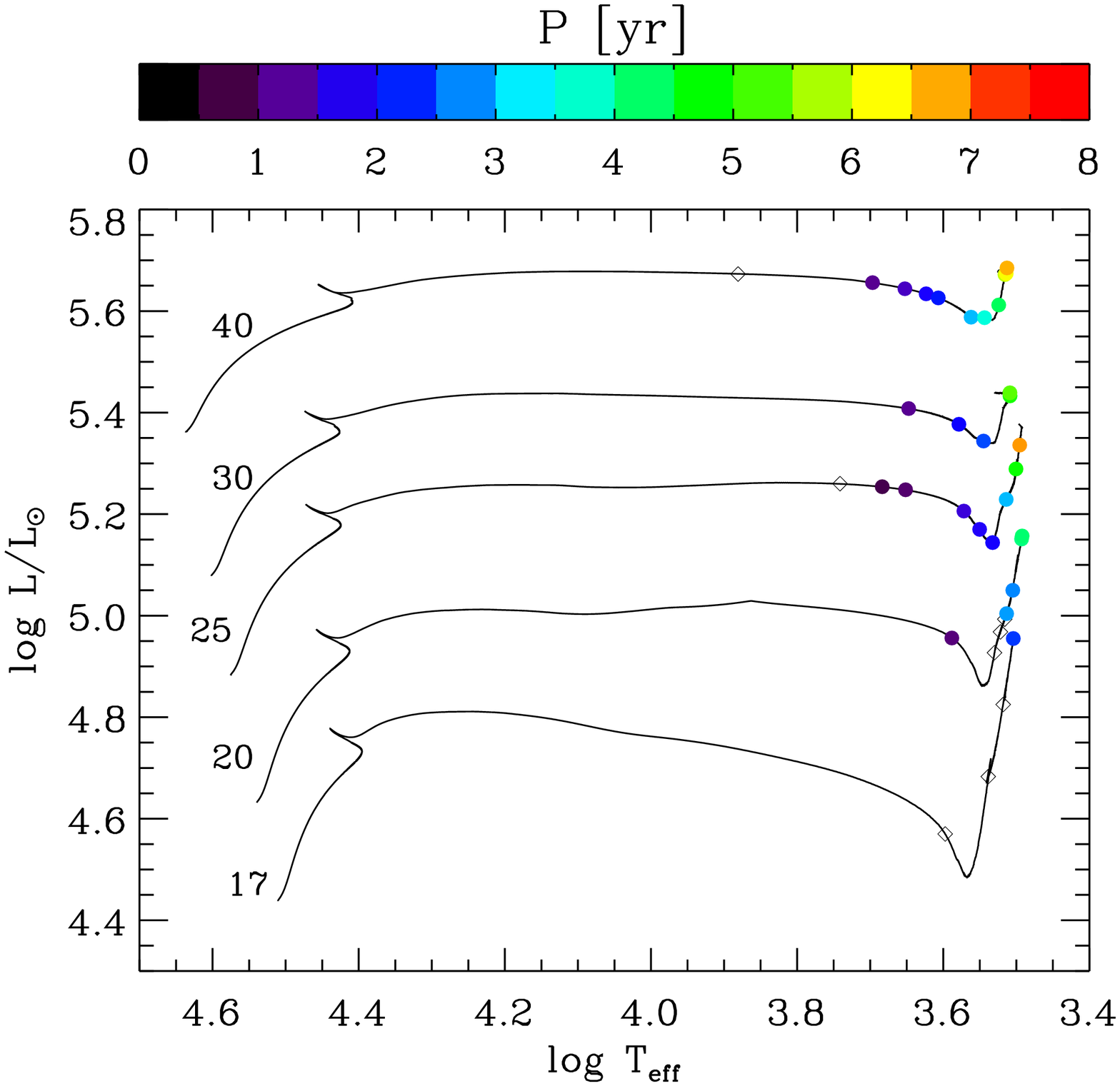} 
\plotone{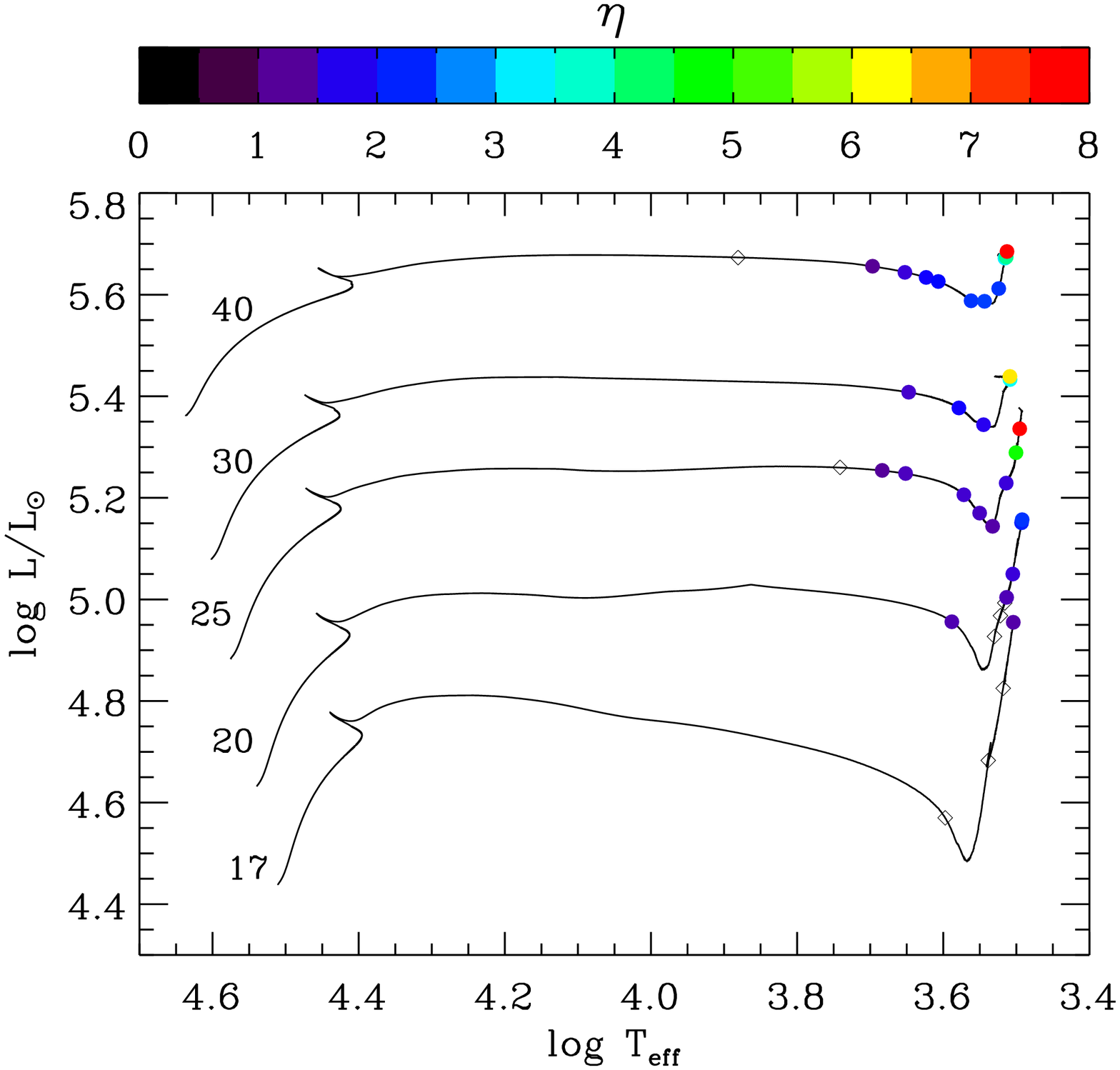} 
\caption{
Evolutionary tracks of hydrostatic reference models with initial masses of 17,
20, 25, 30 and 40~\Msun, on the HR diagram.  The color shading gives pulsation
periods (upper panel) and growth rates (lower panel; see the text for the
definition), obtained with the hydrodynamic calculations that are performed at
the given reference points.  The open diamonds mark the reference points where
no pulsation is detected with our code.  
}\label{fig:hr} \end{figure} 

\begin{figure}
\epsscale{1.0} 
\plotone{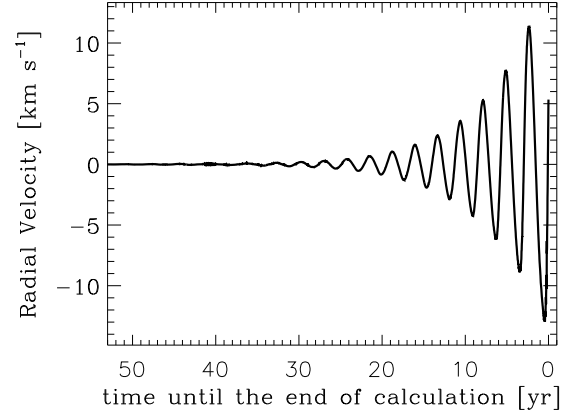} 
\caption{Radial velocity at the surface in a hydrodynamic RSG model sequence 
with  $M_\mathrm{init} =$ 20~\Msun. 
The surface luminosity and the effective temperature at the reference point
are $log L/\mathrm{L_\odot} = 5.05$ and $T_\mathrm{eff} = 3198~\mathrm{K}$, respectively. 
}\label{fig:velocity} 
\end{figure}

To start with, we construct non-rotating hydrostatic models at solar
metallicity for 11 different initial masses (15, 16, 17, 18, 19, 20, 21, 23,
25, 30, 40~\Msun).  Since the hydrodynamic term in the momentum equation is
switched off,  pulsations do not appear in these model sequences. The HR
diagrams for some model sequences are shown in Fig.~\ref{fig:hr}. 

Then, we turn on the hydrodynamic term and restart the calculation from the
reference models chosen at various evolutionary epochs during the supergiant phase 
for each model sequence. In total, 71 reference models are selected
as starting points.   In Fig.~\ref{fig:hr},  some of the starting points are
shown, marked by filled circles and open diamonds.  To guarantee enough time resolution for describing
pulsations in the hydrodynamic calculations,  we limit  the time steps such
that $\Delta t \le
0.01[R/(1000\mathrm{R_\odot})]^2[(15\mathrm{M_\odot})/M]~\mathrm{yr}$.  This
choice is based on the fact that pulsations of a RSG with $ M_\mathrm{init} >
15~\mathrm{M_\odot}$ have a period of the order of 1000 days
\citep[e.g.][]{Heger97}, and obey the relation of $P \propto R^2/M$, where $P$
is the pulsation period, $R$ the radius and $M$ the total mass of a RSG
\citep{Gough65}.

An example of such calculations is given in Fig.~\ref{fig:velocity}.  As the
pulsation amplitude grows gradually, the surface velocity eventually
reaches the sound speed. Since our code cannot describe shock waves, we
investigate the pulsation properties only with the sub-sonic results.
Fig.~\ref{fig:hr} shows the pulsation period $P$ and the growth rate $\eta$ of
the fundamental mode at different reference points on the HR diagram.  Here,
the growth rate is defined as $\eta = |v(t_0+P)/v(t_0)|$, where  $v(t_0)$
denotes a local maximum of the surface velocity located at a certain time
$t=t_0$. According to this definition, pulsational instability means $\eta > 1$.

\begin{figure}
\epsscale{1.0} 
\plotone{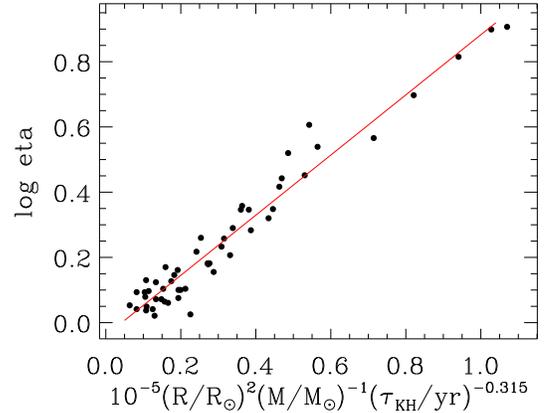} 
\caption{Pulsation growth rate versus some physical quantities of RSGs.  
The solid line gives the best fit, as expressed in  Eq.~(\ref{eq1}).  See the text for more details. 
}\label{fig:eta} 
\end{figure}

The growth rate obtained here should be considered only indicative, given the
limitations of our adopted numerical method.   For example, in our
calculations, no pulsation is detected from some reference points.  This does
not mean that the considered RSGs are stable against the pulsational
instability, but that the growth rate is too small for our code to follow.
Also, in our calculations, instant adjustment of convective flux during
pulsation is assumed, which is not physical given the similar time scales of
convection and pulsation\footnote{Note, however, non-linear calculations by
\citet{Heger97} give quite consistent results with those of their linear
stability analysis that assumes a frozen-in convective flux}. Despite these
uncertainties, our results are qualitatively in good agreement with
those of previous studies: the growth rate is generally higher for a larger luminosity 
to mass ratio ($L/M$), and/or for a smaller thermal time scale of the envelope. Specifically, as
shown in Fig.~\ref{fig:eta}, we find the following relation between the
pulsation growth rate and the stellar structure from our results with the
hydrodynamic model sequences: 
\begin{equation}\label{eq1}  
\log \eta =
\mathrm{C_1}\left(\frac{R}{\mathrm{R_\odot}}\right)^2\left(\frac{M}{\mathrm{M_\odot}}\right)^{-1}%
\left(\frac{\tau_\mathrm{KH,env}}{\mathrm{yr}}\right)^{-0.315} -
\mathrm{C_2}~, 
\end{equation}
where $\tau_\mathrm{KH,env}$ ($:=  GM_\mathrm{tot}M_\mathrm{env}/LR $) denotes
the Kelvin-Helmoltz time scale of the hydrogen-rich envelope, $\mathrm{C_1} =
9.219\times10^{-6}$ and $\mathrm{C_2} = 0.0392844$.

We confirm that RSGs are generally more susceptible to the
pulsation instability for larger initial masses.  Fig.~\ref{fig:hr} indicates
that  for $M_\mathrm{init} > 20~\mathrm{M_\odot}$, pulsations can become strong
enough to be detected with our code even before the star reaches the Hayashi
line, while pulsations are only found after neon burning for $M_\mathrm{init} =
17$~\Msun.  We do not find any signature of pulsation for $M_\mathrm{init} =
15$~\Msun\footnote{\citet{Heger97} found pulsations for the same initial mass.
The reason for this difference is that their models consider rotationally
enhanced mass loss, resulting in a higher $L/M$ than in our non-rotating
models.}. Therefore,  if a super-wind could be induced by strong
pulsations, it would occur in an earlier evolutionary epoch for a higher mass
star.

\section{The evolution of supergiants with super-winds}\label{sect:evol}

\begin{figure}[t]
\epsscale{1.0} 
\plotone{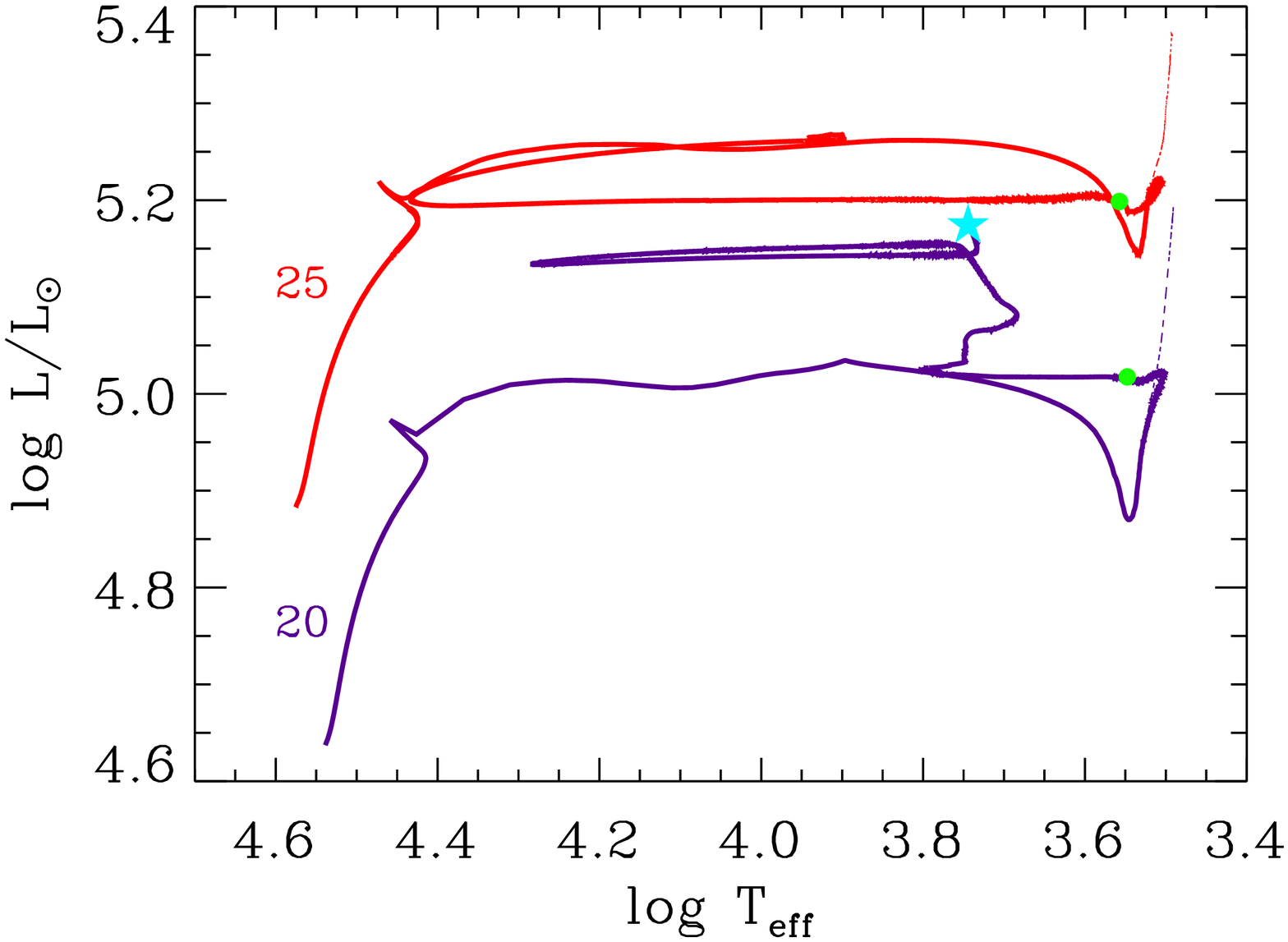} 
\plotone{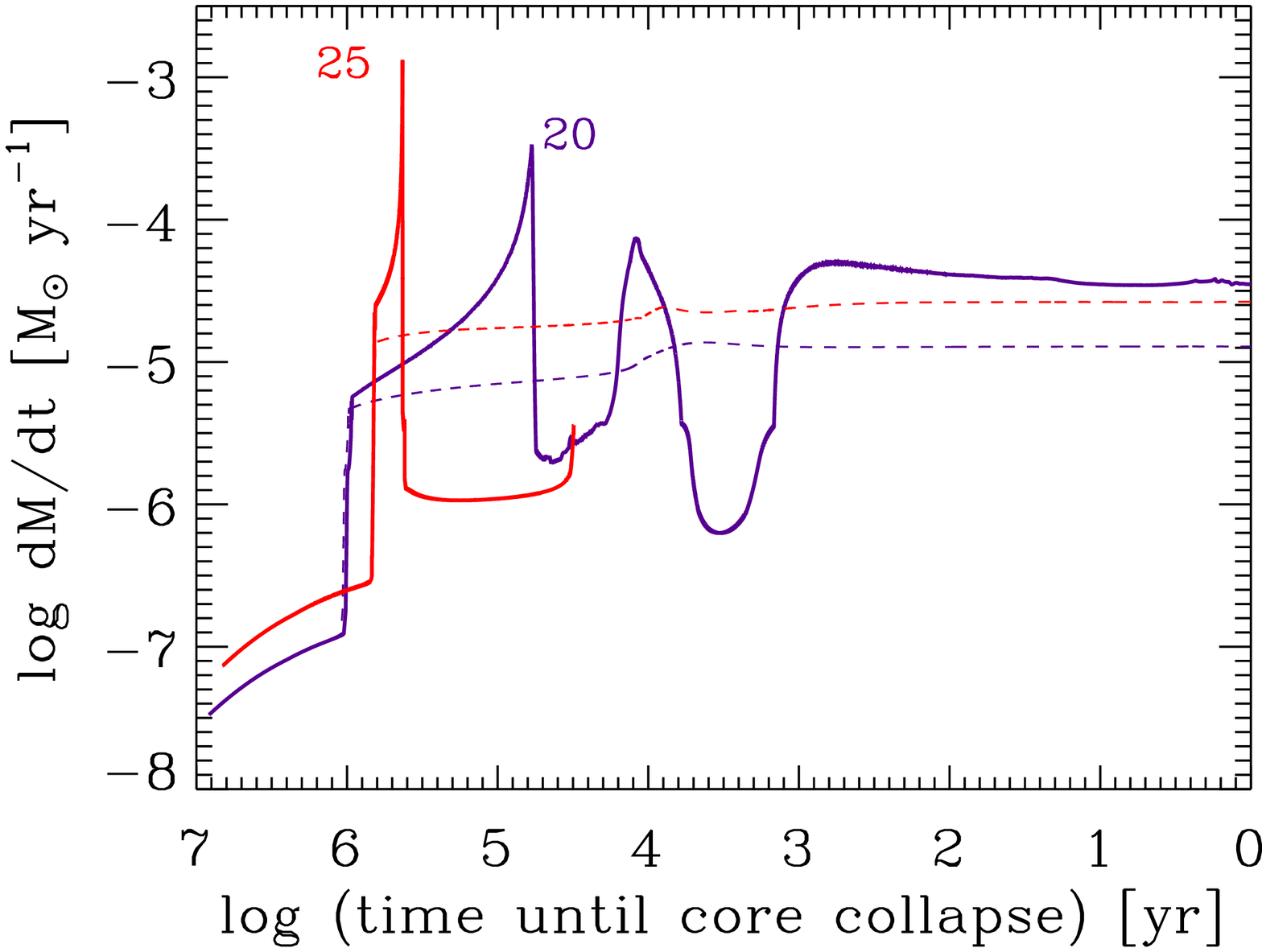} 
\caption{\emph{Upper panel}: 
The evolution of 20 and 25~\Msun{} stellar models on the HR diagram. The thick
solid lines and the thin dashed lines give the evolutionary tracks with
$\alpha=1.75$ (i.e., with pulsation-induced mass loss enhancement), and
$\alpha= 0.0$ (i.e., no enhancement of mass loss)  in Eq.~(\ref{eq2}),
respectively.  The calculation was terminated at a point close to the end of helium
burning (central helium mass fraction equal to 0.0013) for the
25~\Msun{} model sequence with $\alpha=1.75$, while the evolution is followed
up to neon burning for the other sequences.  The filled circle on each
track with $\alpha=1.75$ marks the position when the mass loss rate reaches its
peak.  The filled star gives the position of the pre-supernova star for the
20~\Msun{} sequence  $\alpha=1.75$.  \emph{Lower panel}: The mass loss rates of
the corresponding stellar models.
}\label{fig:evol} 
\end{figure}

If the mass loss rate were dramatically enhanced due to strong pulsations, how
would it affect the evolution of supergiants, and what would be the observational
consequences? 

To address this issue in more detail, we perform an experiment as the
following.  We assume that the mass loss rate of a supergaint is significantly
enhanced compared to the rate of JNH88,  if pulsations are strong enough to be
detected with our hydrodynamic stellar evolution code, i.e. if $\eta > 1$ in
Eq.~(\ref{eq1}): 
\begin{equation}\label{eq2}  
\dot{M} =
\eta^{\alpha}\dot{M}_\mathrm{JNH88}, 
\end{equation}
where $\dot{M}_\mathrm{JNH88}$ is the mass loss rate of JNH88, and $\alpha$ a free
parameter.  New evolutionary model sequences are constructed, with this
modified mass loss rate.  Here, the hydrodynamic term is switched off again,
and normal evolutionary time steps are used.

It should be kept in mind that the empirical mass loss rate of RSGs of JNH88
might already reflect the role of pulsations.  However, as implied by the
growth rates shown in Fig.~\ref{fig:hr}, only very unstable massive stars with
$M_\mathrm{init} \ga 19$~\Msun{} are  significantly affected by this
modification.  Given the paucity of RSGs with $M_\mathrm{init} > $20 --
25~\Msun{} in our galaxy and nearby galaxies \citep[e.g.][]{Massey09}, our
assumption that the mass loss history of such very unstable RSGs can be
qualitatively different from that of most RSGs in the sample of JNH88  does not
necessarily cause an inconsistency.  

Fig.~\ref{fig:evol} shows the evolutionary tracks and the mass loss rates of
model sequences with $\alpha = 1.75$ (this choice for $\alpha$ is rather
arbitrary), compared to the corresponding ones with $\alpha = 0$, for two
different masses ($M_\mathrm{init} =$ 20 and 25~\Msun).  Two important points
are worthy of notice here.  

First, as the star gradually loses mass,  the $L/M$ ratio and the pulsation
growth rate increase accordingly. If the value of $\alpha$  is large enough (as
adopted here), this leads to a runaway increase in the mass loss rate as shown
in the lower panel of the figure. Second, the mass loss rate decreases very
rapidly when the star moves away from the pulsationally unstable regime on the
HR diagram. The peak of the mass loss rate is reached when the helium mass
fraction at the centre decreases to 0.54 (0.031), and the total mass to
9.35~\Msun{} (6.50~\Msun{}), for 25~\Msun{} (20~\Msun{}) model sequence. 
As shown for the 20~\Msun{} sequence, such a mass loss history
of runaway increase followed by sudden decrease can be repeated as the star
moves in and out the unstable regime.  The 20 \Msun{} star finally ends
its life as a yellow supergiant of $M_\mathrm{tot} = 6.1$~\Msun, with only a
small amount of hydrogen of about 0.5~\Msun. The most likely outcome of the
death of such a star would be a Type IIb supernova.  The 25~\Msun{} star has a
hydrogen envelope of 0.22~\Msun{} at core helium exhaustion, and would
produce either a Type Ib or a IIb supernova depending on the subsequent
history of mass loss. 

When the mass loss rate reaches the maximum, the growth rate given by
Eq.~(\ref{eq1}) becomes as high as 11.3 and 10.6 for 25 and 20~\Msun{} models,
respectively.  We checked if non-linear evolutionary calculations also give
such high growth rates, given that the relation of Eq.~(\ref{eq1}) is only
based on the result with $\eta \la 8.0$ (Fig.~\ref{fig:eta}).  We find that, at
the reference points marked by the filled circles in Fig.~\ref{fig:evol}, the
hydrodynamic calculations give $\eta = $ 9.9 and 9.0 for 25 and 20 \Msun{}
models respectively,  which are comparable to the values given by Eq.~(\ref{eq1}).

\section{Discussion}\label{sect:discussion}

If a pulsationally-driven super-wind (PDSW) phase could be induced by strong
pulsations, this would have very important implications for supernova
progenitors.  As implied by our model sequences presented above, such a PDSW
can significantly reduce the critical ZAMS mass ($M_\mathrm{crit}$) above which
a huge fraction of the stellar envelope is removed before core collapse, thus
producing no Type II-P supernova.  Based on our results, we suggest
$M_\mathrm{crit} \sim$ 19...20\Msun{} with a PDSW (see Figs.~\ref{fig:hr} and
\ref{fig:evol}), while the models with the JNH88 mass loss rate predict SN II-P
progenitors up to $\sim$ 25\Msun.  
Interestingly, this value of $M_\mathrm{crit}$
is comparable to what  other alternative  prescriptions of RSG winds mass loss
predict \citep{Salasnich99, vanBeveren07}. 
Note that $M_\mathrm{crit}$ could be even
further reduced by rotation, as implied by the result of \citet{Heger97}~\citep[see also][]{Meynet03}.

This might provide a plausible solution to the so-called RSG problem, i.e. the observed
lack of type II-P progenitors with $M_\mathrm{init} \ga 16.5 \pm 1.5$\Msun
\citep{Smartt09}.  Such observation could also result from the presence of
dusty circumstellar material around the stars, as an obscured progenitor would
be estimated to have a lower initial mass in pre-SN images.  However
\citet{Smartt09} also noted that the  number of SNe type II-P in their sample
is consistent with the expected number of stars in the range 8.5-16.5~\Msun{}
assuming a Salpeter IMF.  This could be a possible indication of the fact that
stars with higher initial mass do actually lose most of their hydrogen
envelope, which cannot be easily understood with the canonical mass loss rate
of JNH88. 

The PDSW  is expected to significantly affect the circumstellar medium around a
RSG (van Veelen, in prep.).  When the star dies, the shock produced by the
collision between the SN ejecta and the circumstellar material transforms
kinetic energy into thermal energy. This energy can be radiated away at
different wavelengths, resulting in brightnening the supernova remnant for long
times \citep{Chevalier77}.  If the collision occurs directly after the SN
explosion, the emission can even alter the spectrum and light curve of the SN.
This scenario is usually invoked to explain the sub-class of Type IIn
supernovae \citep{Schlegel90,Filippenko97}.  To reproduce the light curve of
the most luminous Type IIn supernovae (SNe IIn) like SN 2006gy and SN 2006tf,
very massive shells are required, which need to be ejected in eruptive events
a few years before core collapse \citep[e.g., ][]{vanMarle10}. Pulsational pair
instability \citep{Woosley07} and LBV-like eruptions \citep[e.g., ][]{Smith07}
have been discussed to explain the  presence of shells of 10-25\Msun{} around
the progenitors of luminous SNe IIn.  A PDSW phase is an unlikely explanation
for such extreme environments. However it might be interesting for those type
IIn where the required mass in the stellar vicinity is of the order of a few
solar masses or less.  

In this context it is interesting to consider the circumstellar medium around
the RSG VY CMa.  The stellar surrounding appears shaped by episodic mass
ejections which occurred about 500-1000 yr ago \citep{Smith09}. The complex
morphology of the CSM is suggestive of a possible interaction between
convection and pulsation \citep[see Fig.~13 in][]{Smith09}, which for these
stars are predicted to occur on similar timescales \citep[e.g., ][]{Heger97}.
The mass loss rate of $1-2\times10^{-3}~\mathrm{M_\odot yr^{-1}}$ derived by
\citet{Smith09} is comparable to the one expected during the PDSW phase (see
Fig.~\ref{fig:evol}). Even if the mass loss prescription for PDSW we used is
somewhat arbitrary, we want to stress that the energy available from the growth
of pulsation is enough to drive mass loss rates up to $\sim
10^{-2}~\mathrm{M_\odot yr^{-1}}$. \citet{Smith09} state that an extreme RSG
like VY CMa would produce a Type IIn event like SN 1988Z if it were to explode
in its current state. Therefore the occurrence of pulsation-driven super-winds
in RSG might explain moderately luminous  SN IIn if the enhanced mass loss
takes place less than about $\sim10^3$ years before core collapse.  With the mass loss
prescription of Eq.~\ref{eq2} with a sufficiently large $\alpha$ ($\sim 2$),
this would occur in a narrow range around $\sim$19\Msun{} for non-rotating
stars. An accurate determination of the expected rate of type IIn due to PDSW
requires a more realistic mass loss prescription. 

To conclude, if a pulsationally-driven super-wind phase could be induced by
strong pulsations, we would expect a substantial change in the late evolution of
single massive stars. The mass loss rate during the RSG phase would increase
dramatically for stars with $M_\mathrm{init} \ga M_\mathrm{crit}$ compared to
the JNH88 rate, and the PDSW phase should start earlier in the evolution of
more massive stars. The resulting pre-supernova structure of these stars is
affected, as well as their CSM.  For single stars, this suggests the following
sequence in supernova types as function of increasing initial mass:
II-P~$\xrightarrow{}$~II-L~$\xrightarrow{}$~IIn~$\xrightarrow{}$~IIb~$\xrightarrow{}$~Ib~$\xrightarrow{}$~Ic.
Detailed estimate for the mass range of each supernova requires a better
prescription for the pulsationally enhanced mass loss, which deserves
further study. 

\vskip 0.5cm
We are grateful to Stan Woosley, Bob van Veelen and Norbert Langer for many useful discussions. 
S.C.Y. is supported by the DOE SciDAC Program (DOE DE-FC02-06ER41438).  
\vskip 0.5cm

\end{document}